\documentclass[sn-mathphys,Numbered]{sn-jnl}

\usepackage{graphicx}%
\usepackage{multirow}%
\usepackage{amsmath,amssymb,amsfonts}%
\usepackage{amsthm}%
\usepackage{mathrsfs}%
\usepackage[title]{appendix}%
\usepackage{xcolor}%
\usepackage{textcomp}%
\usepackage{manyfoot}%
\usepackage{booktabs}%
\usepackage{algorithm}%
\usepackage{algorithmicx}%
\usepackage{algpseudocode}%
\usepackage{listings}%
\usepackage{csquotes}
\usepackage{amsmath}
\usepackage{pdflscape}
\usepackage{breqn}
\usepackage{subcaption}
\usepackage{tikz}
\usepackage{csquotes}
\usepackage{tipa}
\usepackage{lineno}

\usepackage{graphicx}
\usepackage{epstopdf, epsfig}
\usepackage{array}
\usepackage{marginnote}
\usepackage{hyperref}

\hypersetup{
    colorlinks=true,    
    linkcolor=black,    
    citecolor=black,    
    urlcolor=blue       
}

\theoremstyle{thmstyleone}%
\theoremstyle{thmstyletwo}%

\theoremstyle{thmstylethree}%
\usepackage{graphicx}
\usepackage{epstopdf, epsfig}
\usepackage{tikz}
\usepackage{dashrule}

\usepackage{amsmath}
\usepackage{flowchart}
\usepackage{caption}
\usepackage{subcaption}
\usepackage[font=small,labelfont=bf]{caption}
\definecolor{ao(english)}{rgb}{0.47,0.67,0.19}
\definecolor{kd-color}{rgb}{0.00,0.45,0.74}
\definecolor{kd+color}{rgb}{0.93,0.69,0.13}
\definecolor{kp-color}{rgb}{0.85,0.33,0.10}
\definecolor{kKH+color}{rgb}{0.49,0.18,0.56}

\begin{document}


\title[Article Title]{Wave reflections and resonance in a Mach 0.9 turbulent jet}


\author*[1]{\fnm{Robin} \sur{Prinja}}\email{robin.prinja@univ-poitiers.fr}

\author[1]{\fnm{Eduardo} \sur{Martini}}\email{eduardo.martini@ensma.fr}

\author[1]{\fnm{Peter} \sur{Jordan}}\email{peter.jordan@univ-poitiers.fr}

\author[2]{\fnm{Aaron} \sur{Towne}}\email{towne@umich.edu}

\author[3]{\fnm{Andr{\'e}} \sur{V. G. Cavalieri}}\email{andre@ita.br}

\affil[1]{D{\'e}partement Fluides, Thermique et Combustion, Institut Pprime, CNRS - Universit{\'e} de Poitiers - ENSMA, Chasseneuil du Poitou, Nouvelle-Aquitaine, France}

\affil[2]{Department of Mechanical Engineering, University of Michigan, Ann Arbor, MI 48109, USA}

\affil[3]{Instituto Tecnol{\'o}gico de Aeron{\'a}utica, S\~{a}o Jos{\'e} dos Campos/SP, Brazil}



\abstract{This work aims to provide a more complete understanding of the resonance mechanisms that occur in turbulent jets at high subsonic Mach number, as shown by   
Towne et al. (\textit{J. Fluid Mech.}, vol. 825, 2017, pp. 1113–1152). Resonance was suggested by that study to exist between 
upstream- and downstream-travelling guided waves. Five possible resonance mechanisms were postulated, each involving different families of guided waves that reflect in the nozzle exit plane and at a number of downstream turning points.
{However, that study did not identify which of the five resonance mechanisms underpin the observed spectral peaks.}
In this work, the waves underpinning resonance are identified via a biorthogonal projection of Large Eddy Simulation data on eigenbases provided by a locally parallel linear stability analysis.
Two of the five scenarios postulated by Towne et al. are thus confirmed to exist in the turbulent jet. 
The reflection-coefficients in the nozzle exit and turning-point planes are, furthermore, identified.
{Such information is required as input for simplified resonance-modelling strategies such as developed in Jordan et al. (\textit{J. Fluid Mech.}, vol. 853, 2018, pp. 333–358)
 for jet-edge resonance, and in 
 Mancinelli et al. (\textit{Exp. Fluids}, vol. 60, 2019, pp. 1–9)
 for supersonic screech.}

}






\maketitle

\section{Introduction}

The mechanisms underpinning the oscillator behaviour in fluid-mechanics problems can be classified as short- or long-range. Short-ranged mechanisms are typically associated with absolute instability \citep{huerre1985absolute}, observed for instance in cold wakes  \citep{huerre1990local} and hot jets \citep{monkewitz1988absolute}. Long-range mechanisms involve a pair of upstream- and downstream-travelling waves that interact at two end locations, where they are reflected into one another. If the wave amplitude increases over the cycle between two reflections, a long-range-resonant instability occurs. If the amplitude is unchanged, a neutrally stable global mode is created, which, in turbulent flows, can be driven by the background turbulence. Such mechanisms have been observed in many different flows, such as when jets interact with edges \citep{powell1953edge, jordan2018jet}, in cavity flows \citep{rockwell1979self,rowley2002self}, impinging jets \citep{ho1981dynamics,tam1990theoretical},  shock-containing jets \citep{raman1999supersonic,edgington2019aeroacoustic}, and high subsonic jets \citep{towne2017acoustic,schmidt2017wavepackets}.
The waves underpinning resonance can frequently be modelled using linear mean-flow analysis \citep{michalke1970note,crighton1976stability,jordan2013wave,cavalieri2019wave}.

In this study, we revisit the tones found in a turbulent jet {with a Mach number of $M=0.9$}, postulated by \citet{towne2017acoustic} to be driven by waves resonating between the nozzle exit and downstream turning points. 
{Turning points represent spatial locations where the upstream- and downstream-travelling waves
can interact and exchange energy through reflection and transmission processes.
The turning point is a downstream location characterized by the presence of a saddle point where a pair of upstream- and downstream-travelling waves share the same frequency, wavenumber and phase velocity.
Upstream of this point, the downstream travelling wave is propagating, while  becoming evanescent after it. As evanescent waves do not propagate energy, the energy of the incident wave is typically transferred to an upstream-travelling reflected wave \cite{hirschberg2004introduction}}.

The waves in question are guided waves of positive and negative generalised group velocities, denoted as $k^+$ and $k^-$ respectively, as per \citet{briggs1964electron} and \citet{bers1983space}. 
These guided waves are neutrally stable at the resonance frequencies \citep{towne2017acoustic, martini2019acoustic}  and can be described using locally parallel linear stability analysis.
The waves consist of one downstream-travelling wave: the $k^+$ duct-like mode{, denoted here as $k^+_T$,} and two upstream-travelling waves: the $k^-$ duct-like mode and $k^-$ {discrete shear-layer  mode, denoted here as $k^-_d$ and $k^-_p$ respectively}. 
Additionally, the Kelvin-Helmhotz (hereafter K-H) mode \citep{michalke1970note}{, denoted here as $k^+_{KH}$}, 
which characterizes the convective instability of the jet, is also considered. 
However, the K-H mode is not the primary focus of our study as it does not contribute to the resonance mechanisms under investigation.

\begin{figure}
    \centering
    \includegraphics[width=0.6\textwidth]{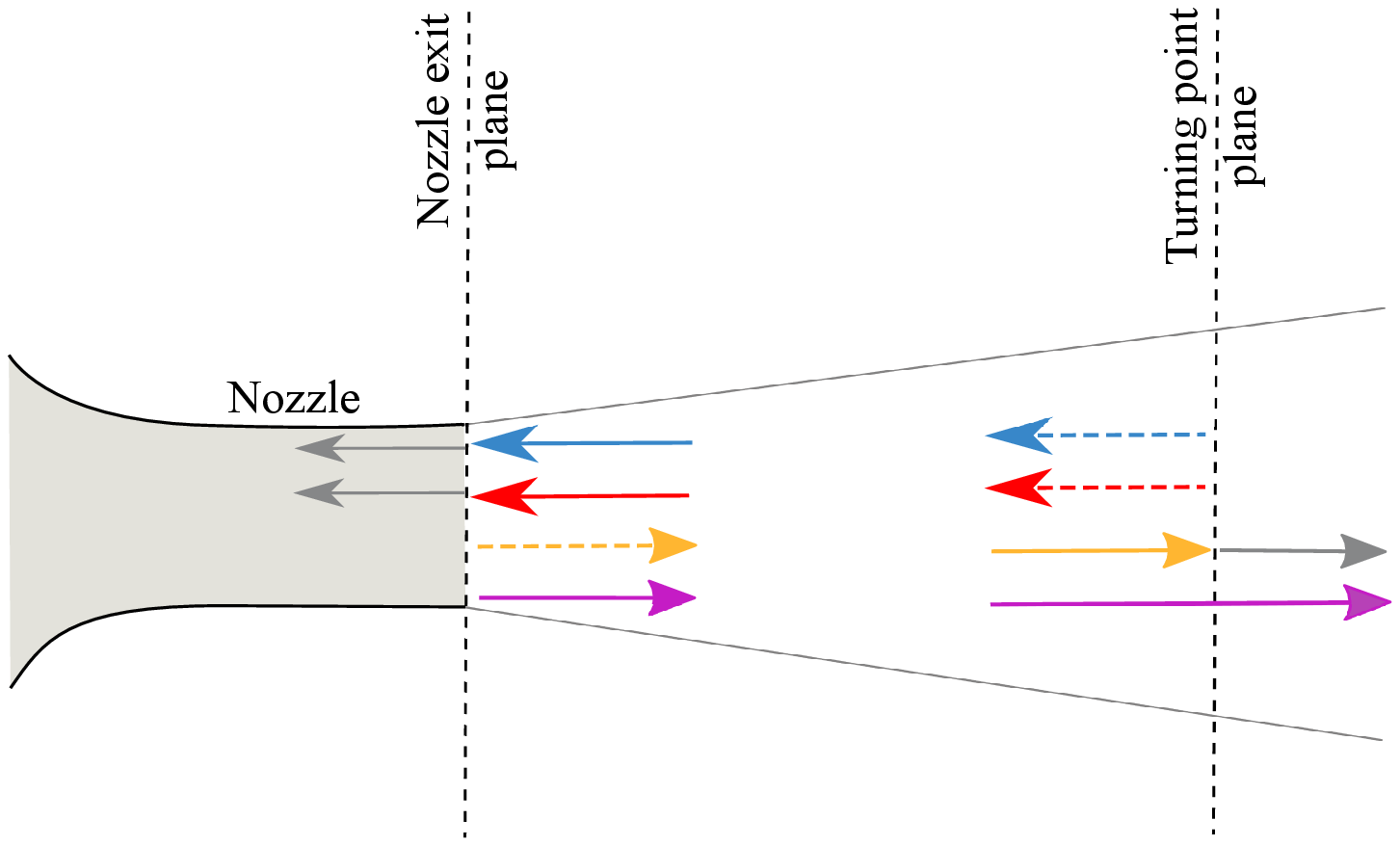}
   \caption{Sketch of waves interacting at the resonance end locations. At the nozzle exit plane: ($\longleftarrow$, blue) incident $k^-_d$ wave; ($\longleftarrow$,  red) incident $k^-_p$ wave; ($\longleftarrow$,  grey) transmitted waves; ($\longrightarrow$, dashed yellow) reflected $k^+_T$ wave; ($\longrightarrow$,  purple) $k^+_{KH}$ wave. At the turning point plane: ($\longrightarrow$,  yellow) incident $k^+_T$ wave;  ($\longrightarrow$,  grey) transmitted wave;  ($\longleftarrow$, dashed blue) reflected $k^-_d$ wave;  ($\longleftarrow$, dashed red) reflected $k^-_p$ wave.}
   \label{fig:nozzle}
\end{figure}

{
Figure \ref{fig:nozzle}  shows a schematic depiction  of reflections at the nozzle exit plane and at a downstream turning-point plane, the spatial position of which depends on the frequency. 
At resonant frequencies, the $k^+$ and $k^-$ guided waves propagate between the nozzle exit and turning point, exchanging energy through 
reflections at these end locations. 
At the turning point, the incident propagative $k^+_T$ wave can be reflected as a propagative wave ($k^-_d$ or $k^-_p$) and transmitted as an evanescent wave. The reflected wave then propagates upstream until it reaches the nozzle exit plane where it is reflected as a $k^+_T$ wave that travels downstream until the turning point, hence completing the resonance loop.
}

Depending on the frequency, the $k^+_T$ wave can form a turning point with either of the $k^-$ waves, resulting in two possible resonance mechanisms. 
These two resonance mechanisms are among five possible mechanisms that potentially exist in the flow, as proposed by \citet{towne2017acoustic}. { While there were indications and speculation in that work regarding the active resonance mechanisms, the lack of reflection-coefficient data prevented a definitive conclusion. Consequently, } it remained unclear which mechanisms were actually active in these turbulent jet flows and why. { In this study, we address this question and provide a conclusive answer.} By thoroughly examining the time-resolved turbulent jet data, we quantitatively evaluate the presence of these resonance mechanisms within the specified tonal frequency range.

We thus revisit the turbulent jet data with the goal of: (1) educing the waves present in the data; (2) establishing which of these underpin resonance; (3) computing the reflection-coefficients associated with energy exchange at the resonance end locations. 

This third objective is crucial for simplified resonance models, such as proposed by \citet{jordan2018jet,mancinelli2019screech,mancinelli2021complex}, where reflection coefficients serve as essential components. 
{
In such models, the conditions required for resonance to occur for a pair of $k^+$ and $k^-$ waves involve both magnitude and phase constraints, respectively,}

\vspace{-0.4cm}
{
\begin{align}
e^{\Delta \alpha_i L} &=\mid R_1 R_2\mid, \\
\Delta \alpha_r L + \phi &= 2n\pi,
\end{align}
where \( R_1 \) and \( R_2 \) are the complex reflection coefficients at the resonance end locations, \(\Delta \alpha = \Delta \alpha_r + i \Delta \alpha_i\) represents the difference between the complex axial wavenumbers for \(k^+\) and \(k^-\) waves, $i$ is the imaginary unit, 
\(\phi\) is the argument of \( R_1 R_2 \), 
$L$ is the distance between the two resonance end locations, 
and \( n \) is an integer.
 In the existing models, the magnitudes and phases of reflection coefficients, $R_1$ and $R_2$, are unknown and thus amount to parameters with which the models may tuned to match data, rather than being informed based on flow physics.
The physical representivity of such models is enhanced by the inclusion of data- or flow-physics-based reflection coefficients.}

The paper is organised as follows. Section \ref{sec:les_processing} presents the Large Eddy Simulation (LES) database which is used. Local linear stability analysis is performed on the jet mean flow in section \ref{sec:stability}. The LES data is then decomposed using bi-orthogonal projections on the stability eigenbasis in section \ref{sec:biorthogonal}.
It is shown how, at resonant conditions, the LES data can be represented by a rank-4 model.
This is the basis for the calculation of 
reflection-coefficients at the resonance end locations.   Section \ref{sec:reflection_method} presents the reflection-coefficient eduction methodology and section \ref{sec:results} presents the final results for a range of resonant frequencies. 

\section{LES database} \label{sec:les_processing}


We analyse LES data for an isothermal jet  {with a Mach number of $M=0.9$} from \citet{bres2018importance}.
The jet is issued from a convergent-straight nozzle  at a  diameter-based Reynolds number of $1 \times 10^6$.
The nozzle-exit boundary layers are fully turbulent.
For this jet flow, the guided waves have been observed in the potential-core region and associated discrete spectral tones have been detected in the near-nozzle region \citep{towne2017acoustic, schmidt2017wavepackets, bogey2021acoustic}. 

The data, described in \citet{bres2018importance}, covers a cylindrical grid with length $30D$ and radius $6D$, where $D$ is the jet diameter. It contains $10000$ timesteps over $2000$ acoustic time units ($tc/D$, where $c$ is sound speed), sampled every $0.2$ acoustic time units. The cylindrical coordinate system has its origin centered on the jet axis in the nozzle plane.

{LES fluctuation time-series data is represented by the vector $ \mathbf{q}_{\text{LES}}  = \left[
\begin{array}{ccccc}
   {\rho} & {u}_x & {u}_r & {u}_{\theta} &  {T}
\end{array}  \right]^\top $, where $\top$ represents the  transpose, ${\rho}$ the density, ${u}_x$ the streamwise velocity, ${u}_r$ the radial velocity, ${u}_{\theta}$ the azimuthal velocity and ${T}$ the temperature.}
This vector is decomposed into Fourier modes,

\begin{equation} \label{eq:lesDecomposition}
\textbf{q}_{\text{LES}} (x,r,\theta,t)= \sum_{\omega} \sum_{m} \mathbf{\hat{q}}_{\text{LES}} (x,r,m,\omega) \hspace{1.3mm} e^{im\theta} e^{i\omega t},
\end{equation}
\noindent where  $x$ is the axial coordinate, $r$ is the radial coordinate,  $\theta$ is azimuthal coordinate, $m$ is the azimuthal wavenumber and $\omega$ is the angular frequency of fluctuation quantities. The time-series is split into 153 realisations, where each realisation contains 256 snapshots and an overlap of 75$\%$. This leads to the frequency resolution of $\Delta St=\Delta f D/ U_{j}=0.0217$, where Strouhal number, $St= f D/ U_{j}$, $f$ is the frequency of fluctuations and $U_{j}$ is the jet velocity. 

\section{Decomposing turbulent jet data into the resonating modes} \label{sec:decomposingField}
To identify the waves that dominate the jet dynamics at the resonant frequencies, the LES fluctuation data at a given streamwise station is projected onto eigenmodes obtained from a locally parallel linear stability analysis that is described in the following section.

\subsection{Local Stability Analysis} \label{sec:stability}
Stability analysis is performed around the LES turbulent mean flow. 
The radial profile of the LES mean flow is fitted with an analytical profile \cite{michalke1984survey}
\begin{equation} \label{eq:mean_flow_fit}
\overline{U}_x(r)= \frac{U_j}{2} \bigg[ 1+ tanh \bigg\{ b \bigg( \frac{0.5}{r/D} - \frac{r/D}{0.5} \bigg) \bigg\}\bigg],
\end{equation}
where $b$ is the fitting parameter.
{The fitting process is performed to ensure a smooth flow profile so as to avoid any spurious or unphysical modes during stability analysis.}
This is done so as to prevent errors that may arise in using the LES mean-flow data: as this is stored on a different grid to that used for the stability analysis, it must be interpolated on to the Chebyshev grid necessary for the pseudo-spectral approach used for the stability analysis; such interpolation can lead to error, particularly following application of the differentiation operators as occurs in the eigenvalue problem described in what follows.

\begin{figure}
    \centering
    \includegraphics[width=0.5\textwidth]{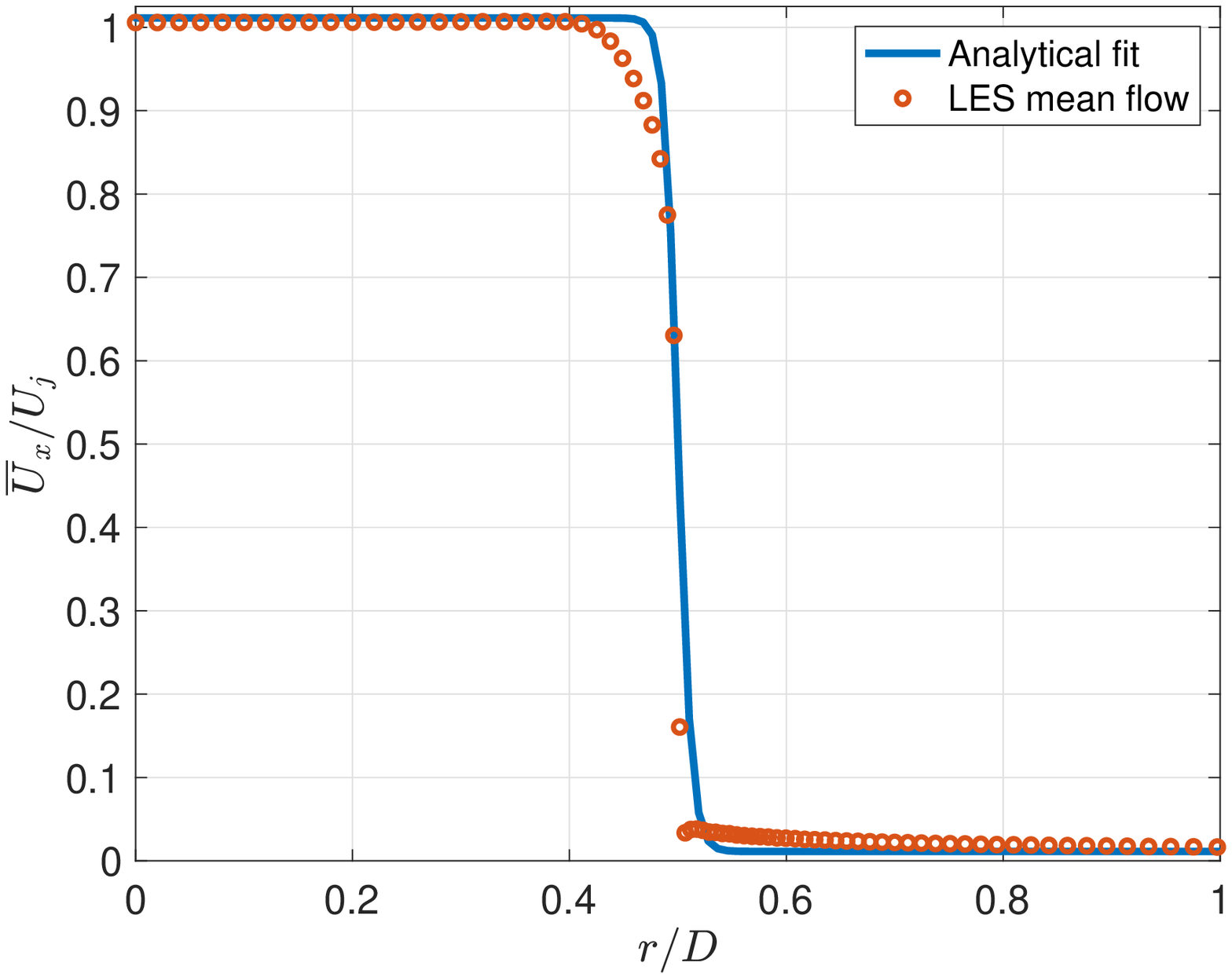}
    \caption{Streamwise velocity mean profile at $x/D=0$ for $M=0.9$ jet:
    Fitted mean flow vs mean flow extracted from LES}
        \label{fig:mean_statistics}
\end{figure}

The radial profiles of the fitted mean flow and the mean flow extracted from LES at $x/D = 0$ are shown in figure \ref{fig:mean_statistics}.
Similarly, fitted mean flow profiles were obtained at other streamwise locations to perform the corresponding local stability analysis.

The normal-mode ansatz for the fluctuation field is given as,
\begin{equation} \label{eq:fluctuation_form}
\mathbf{q}' (x, r, \theta,t) = \mathbf{\hat{q}}(r) e^{i \alpha x} e^{i m \theta} e^{-i \omega t},
\end{equation}
\noindent 
where $ \mathbf{q}' = \left[
\begin{array}{ccccc}
   {\rho}' & {u}_x' & {u}_r' & {u}_{\theta}' &  {T}'
\end{array}  \right]^\top $ is the vector describing the fluctuating quantities, $\mathbf{\hat{q}}{=\left[
\begin{array}{ccccc}
   {\hat{\rho}} & {\hat{u}}_x& {\hat{u}}_r & {\hat{u}}_{\theta} &  {\hat{T}}
\end{array}  \right]^\top} $
gives the radial structure and $\alpha$ is the {
complex streamwise wavenumber. 
$\alpha$ contains both the phase-speed information ($\omega/\alpha_r$, where $\alpha_r$ is the real part of $\alpha$) and the growth-rate information ($\alpha_i$, the imaginary part of $\alpha$)
 }. 
 
This formulation allows the linearised N-S equations to be compactly written as, 
\begin{equation} \label{eq:LNS_system}
\mathbf{M} \mathbf{\hat{q}} = i \alpha  \mathbf{\hat{q}},
\end{equation}
{with the spatial support, $\mathbf{\hat{q}}(r)$, and streamwise wavenumber, $\alpha$, of the waves obtained via this eigenvalue problem.}

The eigenfunctions, $\mathbf{\hat{q}}(r)$, are then normalised such that each mode has: (1) $0^o$ phase angle for the streamwise velocity fluctuation at the jet axis ($\angle {\hat{u}}_x=0^o$ at $r=0$); and (2) unit Energy norm, $E$ \citep{chu1965energy}, defined as
    \begin{equation} \label{eq:energy norm}
 \mathrm{E} = \int_{0}^{\infty} \Bigg[  
   \frac{\overline{T}}{{{\kappa}} \overline{\rho}M^2} \mid {\hat{\rho}} \mid^2+
 \overline{\rho} \mid {\hat{u}}_x \mid^2  +  \overline{\rho} \mid  {\hat{u}}_r \mid^2  +  \overline{\rho} \mid  {\hat{u}}_{\theta} \mid^2 
 +
 \frac{\overline{\rho}}{{{\kappa}}({{\kappa}} -1)\overline{T}M^2} \mid {\hat{T}} \mid^2
  \Bigg] r \hspace{1mm} dr,
 \end{equation}
{where $\kappa=1.4$ is the specific heat ratio, $\overline{T}$ is the mean temperature and $\overline{\rho}$ is the mean density.}

\begin{figure}
    \centering
\begin{subfigure}{0.49\textwidth}
    \includegraphics[width=1\textwidth]{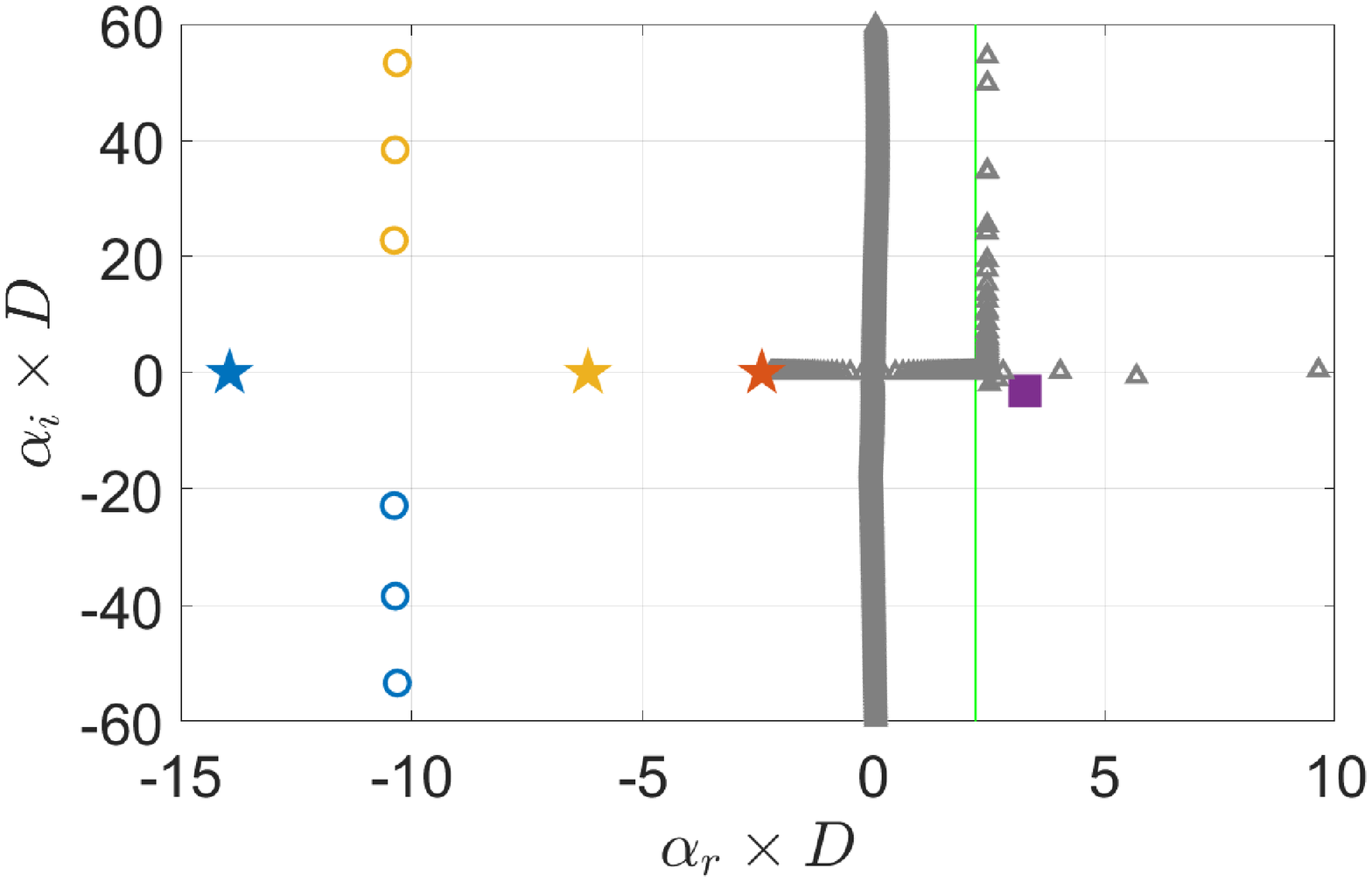}
\vspace{1mm}
    \caption{ }
        \label{fig:eigenspectrum_st03906}
\end{subfigure}   
\hspace{2mm}
\begin{subfigure}{0.45\textwidth}
\hspace{2mm}
    \includegraphics[width=1\textwidth]{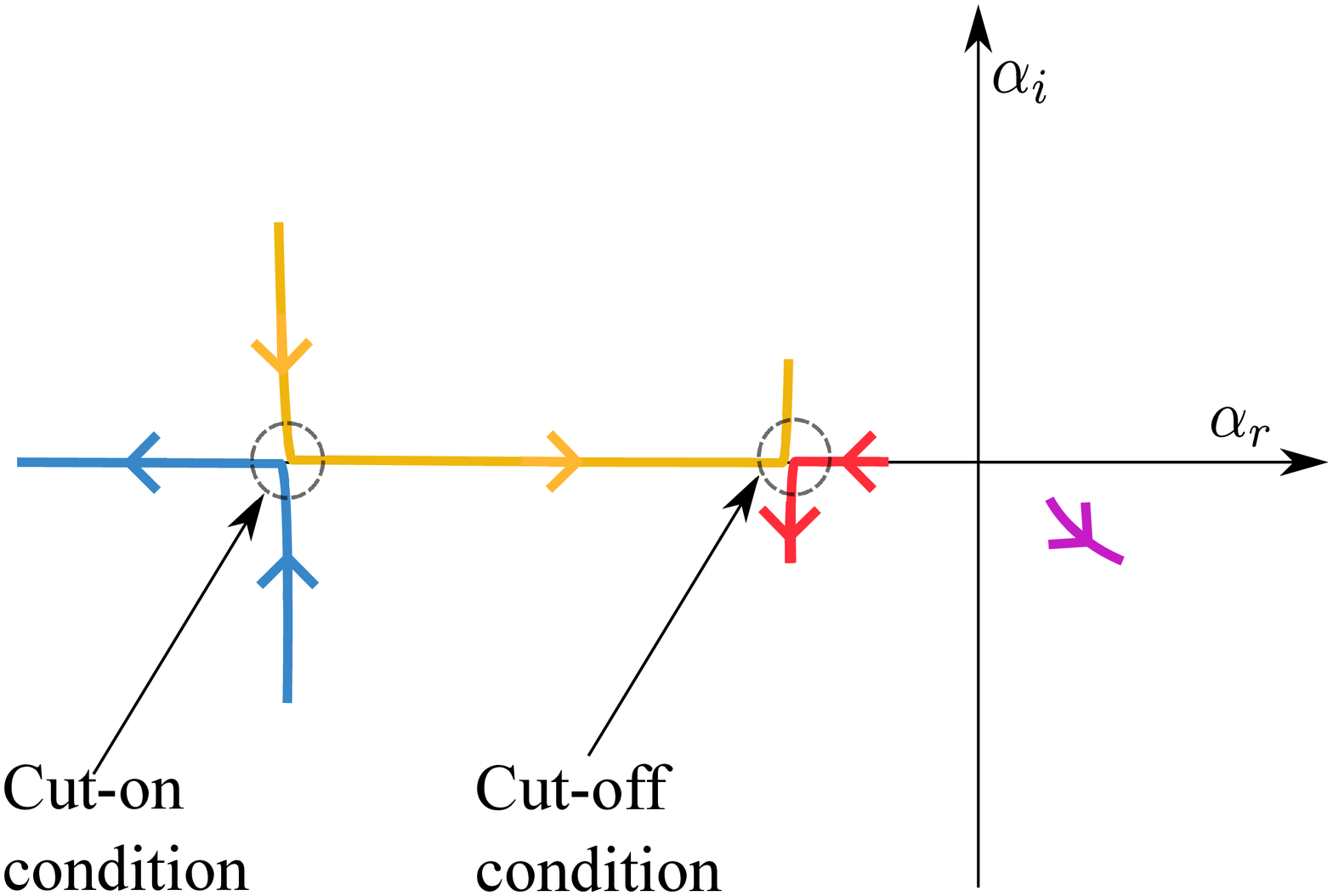}
\vspace{1mm}
    \caption{ }
    \label{fig:trajectory_withSt}
\end{subfigure}  
\caption{Linear stability analysis at $x/D=0$, $m=0$ for $M=0.9$ jet. (a) Eigenspectrum for $St=0.39$: $({\star}$, blue) $k^-_{d}$; $({\star}$, red) $k^-_{p}$; $({\star}$, yellow) $k^+_{T}$; $({\blacksquare}$, purple) $k^+_{KH}$;  (\rule[.45ex]{1.3em}{1pt}, light green) sonic line; 
    {     $({\circ}$, blue) upstream-travelling 
    trapped acoustic waves; 
$({\circ}$, yellow) downstream-travelling 
    trapped acoustic waves;
    $({\triangle}$, grey, horizontal branch) stable critical layer modes; 
    $({\triangle}$, grey, vertical branch) stable modes with support in the core region.} (b) Modes trajectories for $St$=\{$0.23$\hspace{1mm}$\rightarrow$\hspace{1mm}$0.47$\}:  (\rule[.45ex]{1.3em}{1pt}, blue) $k^-_{d}$; (\rule[.45ex]{1.3em}{1pt}, red) $k^-_{p}$; (\rule[.45ex]{1.3em}{1pt}, yellow) $k^+_{T}$; (\rule[.45ex]{1.3em}{1pt}, purple) $k^+_{KH}$.}
        \label{fig:eigenspectrum_trajectories}
\end{figure}

In the present work, we focus on the azimuthal mode $m=0$ and the frequency range $0.23 \leq St \leq 0.47${, as this has been associated with the strongest power spectral density (PSD) peaks in the near-nozzle region associated with potential-core resonance \citep{bres2018importance}. However, the methodology presented here can be easily extended to higher azimuthal modes and other frequency ranges}. 
Stability analysis is conducted here for $m=0$ within this frequency range. The eigenspectrum for one of the tonal frequencies, $St=0.39$, is shown in figure \ref{fig:eigenspectrum_trajectories}(a), where the real and imaginary parts of $\alpha$ are represented on the horizontal and vertical axes, respectively.
 
Various families of modes can be seen in figure \ref{fig:eigenspectrum_trajectories}(a): (i) 
The resonating modes leading to tones are marked with stars and are named as per \citet{jordan2018jet}. 
They are guided propagative modes resonating between the end locations and are the focus of the present work. 
The resonance loop consists of a downstream travelling mode ($k^+_{T}$) and an upstream travelling mode ($k^-_{d}$ and/or $k^-_{p}$). 
    { 
The physics of these modes vary within the tonal frequency range. Modes $k^+_{T}$ and $k^-_{d}$ belong to families of an infinite number of such modes (marked by circles). At low frequencies, the $k^+_{T}$ mode is largely trapped within and guided by the jet, behaving like an acoustic wave in a soft-walled cylindrical duct, while at high frequencies, it gains support in the shear layer and behaves as a shear-layer mode. 
The $k^-_{d}$ mode is trapped within and guided by the jet, acting like a propagative acoustic mode in a soft-walled duct at high frequencies,  
but becomes evanescent below a specific frequency. 
The $k^-_{p}$ mode, however, is primarily a shear-layer mode that becomes evanescent above a certain frequency.} 
Further details about these modes can be found in \citet{towne2017acoustic, schmidt2017wavepackets, martini2019acoustic, jordan2018jet}; 
(ii)
The $k^+_{KH}$ (K-H mode) is marked with a purple square and is the only unstable mode of the system which leads to amplitude growth of the coherent part of the fluctuation field which then stabilises and decays, forming a wavepacket
\citep{jordan2013wave};
(iii) The modes marked with grey triangles in the first quadrant with subsonic phase speed (the sonic line is at $\alpha_r = \omega /c$) are stable and are distributed in two separate branches: a near-horizontal branch consisting of critical layer modes that have support in the shear layer, and a near-vertical branch with eigenfunctions that have support in the core region of the jet \citep{rodriguez2015study}.

Although all guided modes (marked with stars) are propagative at $St=0.39$, this is not the case for all $St$. 
Figure \ref{fig:eigenspectrum_trajectories}(b) shows the trajectories of the three guided modes and the K-H mode in the complex $\alpha-$plane as $St$ varies from $0.23$ to $0.47$. 
At $St=0.23$, $k^+_{T}$ and $k^-_{d}$ are evanescent, and they move gradually towards the $\alpha_i=0$ axis and overlap at saddle-point 1, defining a cut-on condition at $St=0.37$ \citep{towne2017acoustic}. The modes remain propagative until saddle-point 2, a cut-off condition, which occurs at $0.428$, where $k^+_{T}$ and $k^-_{p}$ modes meet. For $St>0.428$, the $k^+_{T}$ and $k^-_{p}$ modes become evanescent. 
At higher frequencies, other modes from the families of $k^+_{T}$ and $k^-_{d}$ cut on leading to resonance, but these scenarios are not considered in the present work since the most energetic resonance occurs for the considered scenario.

\begin{figure}
    \centering
    \includegraphics[width=0.55\textwidth]{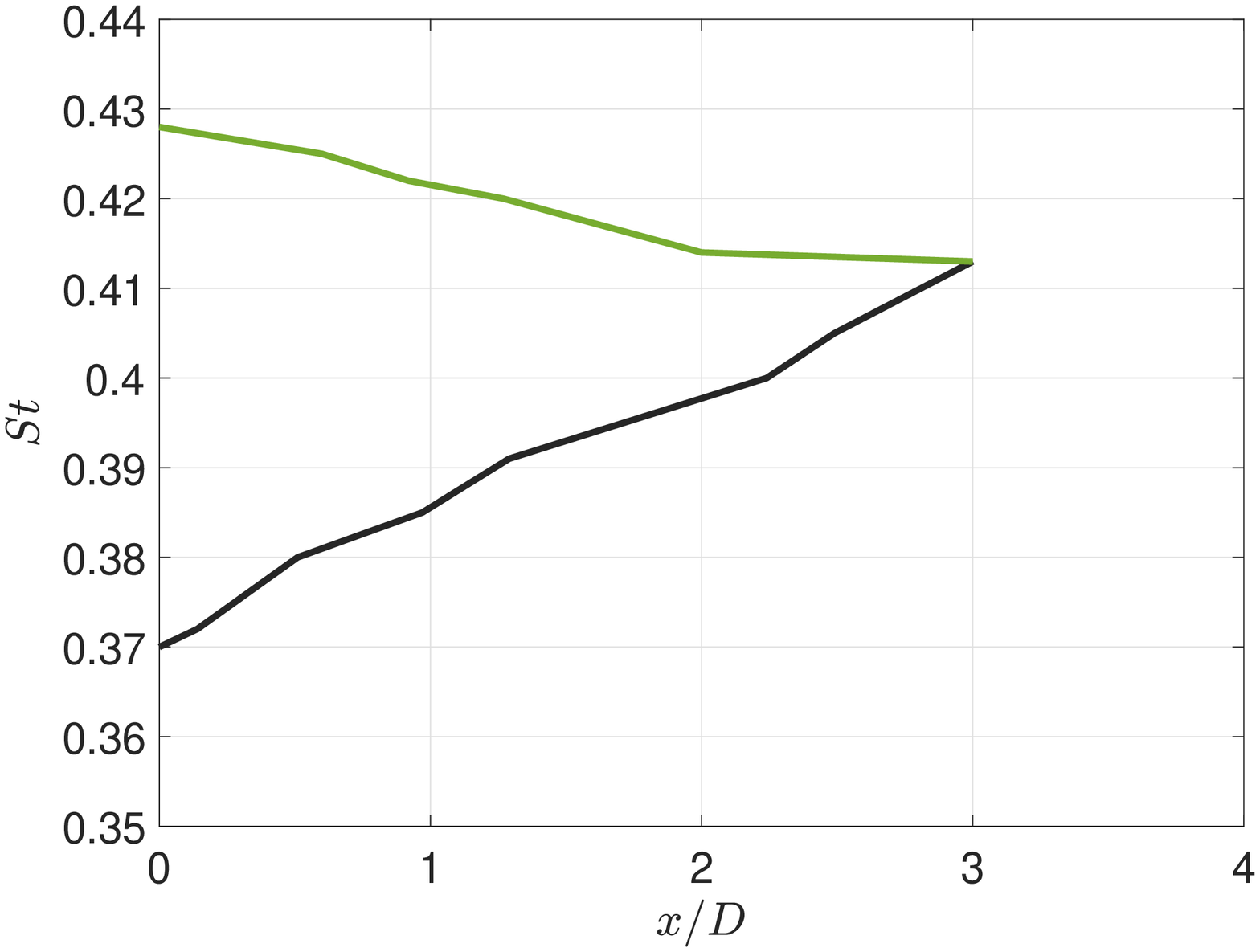}
\caption{Saddle points for $k^-/k^+$ modes from local linear stability analyses: (\rule[.45ex]{1.3em}{1pt}, black) saddle point for downstream- and upstream-travelling duct modes  { i.e. $k^+_{T}$ and $k^-_{d}$}; (\rule[.45ex]{1.3em}{1pt}, green) saddle point for downstream-travelling duct mode and upstream-travelling {discrete shear-layer  mode  i.e. $k^+_{T}$ and $k^-_{p}$}. Strouhal number, $St= f D/ U_{j}$, is plotted against downstream axial distance, $x/D$, from the nozzle exit.}
\label{fig:resonance_mechanism1}
\end{figure}

    { 
    Cut-on and cut-off frequencies are points in the frequency spectrum where mode behavior changes, marking the resonance frequency range.
    These are the saddle point locations corresponding to zero group velocity for the modes where they can interact and exchange energy. 
    At the cut-on frequency, initially evanescent modes become propagative, initiating wave propagation while at the cut-off frequency, modes that were propagative become evanescent.}
For various streamwise locations downstream of the nozzle exit plane i.e. for $x/D>0$, stability analysis is performed to assess the cut-on and cut-off conditions. The results are depicted in figure \ref{fig:resonance_mechanism1} where the frequencies of the two saddle points are plotted as a function of the axial position, as obtained through linear stability analysis in the tonal frequency range $0.37< St < 0.43$.

For any specific tonal frequency within this range, the $k^+_{T}$ wave propagates downstream until it reaches the turning point, the spatial location where $k^+_{T}$ forms a saddle point with either of the $k^-$ waves, depending on the frequency. For $0.37 < St \leq 0.413$ {(frequency range denoted as F1 in this work)}, it is the $k^-_{d}$ wave, while for $0.413 \leq St < 0.428$ {(frequency range denoted as F2 in this work)}, it is the $k^-_{p}$ wave. These represent two distinct resonance mechanisms.

{
The two lines in figure \ref{fig:resonance_mechanism1} show the streamwise dependence of the frequencies of the two saddle points. 
The streamwise position at which they intersect indicates a double saddle point. 
This occurs at $St=0.413$, indicating the possibility of both a double saddle-point ringing mechanism, and a double turning point.
This is possible on account of the fact that, for this frequency, the three modes have the same wavenumbers, the same frequency and the same zero group velocity, allowing the $k^+_{T}$ mode to reflect into and resonate with both $k^-_{d}$ and $k^-_{p}$ modes. 
}



\subsection{Educing mode amplitudes by bi-othogonal projections}  \label{sec:biorthogonal}

We here aim to educe the amplitudes of the three guided waves, which are postulated to be responsible for the observed tones and the K-H mode, which is the main instability of the jet.
These amplitudes when multiplied with their corresponding  eigenfunctions will give the calibrated eigenfunctions for the modes.

The non-normality of the linearized Navier-Stokes equations prevents the eigenfunctions from forming an orthogonal system. However, by leveraging the bi-orthogonality between the eigenfunctions of the direct and adjoint linear stability problems \citep{salwen1981continuous, hill1995adjoint, tumin2007three}, mode amplitudes can be obtained by bi-orthogonal projection following the approach outlined by  \citet{rodriguez2013inlet,rodriguez2015study}. 

A basis for bi-orthogonal projection is constructed from the adjoint system,
\begin{equation} \label{eq:adjoint_system}
\mathbf{M} ^H  \mathbf{\hat{q}}^+ = i \alpha^+ \mathbf{\hat{q}}^+,
\end{equation}
where 
$H$ represents the Hermitian transpose,
$\alpha^+$ are the complex conjugate of eigenvalues of the direct system and $\mathbf{\hat{q}}^+$ are the adjoint eigenfunctions that we seek.

The adjoint eigenfunctions are normalized such that,
\begin{equation}
    (\mathbf{\hat{q}}^+_{{j} })^H   \mathbf{\hat{q}}_{{j} } =1,
\end{equation}
where {$j$} is the index of the mode being normalized.

Before projection, $\mathbf{\hat{q}}_{LES}$ (from \eqref{eq:lesDecomposition}) is interpolated onto the Chebyshev nodes, on which the eigenfunctions are defined, using Piecewise Cubic Hermite Interpolating Polynomials. 
Due to the fast decay of disturbances away from the jet,  the points outside the available LES grid locations are assigned a value of $0$ for the fluctuation quantities. This is corroborated by verifying that $\mid (\mathbf{\hat{q}}^+_{{j}})^H \hspace{1mm}  \mathbf{\hat{q}}_{{j}}  \mid$ is almost the same with or without this assumption $\forall St,  \forall {{j}} $, and hence the projection amplitudes would be negligibly affected.

The mode amplitudes are then obtained by biorthogonal projection,
\begin{equation} \label{eq:biorthogonal_projection}
a_{{j}}^{n} = {(\mathbf{\hat{q}}^+_{{j}})^H \hspace{0.1mm} \hspace{1mm} \mathbf{\hat{q}}_{LES}^{n}},
\end{equation}
where $\mathbf{\hat{q}}_{LES}^{n}$ is the LES fluctuation data from the $n^{th}$ realisation; and 
$a_{{j}}^{n}$ is the expansion coefficient that defines the contribution of the ${{j}}^{th}$ mode to the flow state in the $n^{th}$ realisation, giving the amplitude and the phase of mode.

\begin{figure}
    \centering
\begin{subfigure}{0.49\textwidth}
    \includegraphics[width=\textwidth]{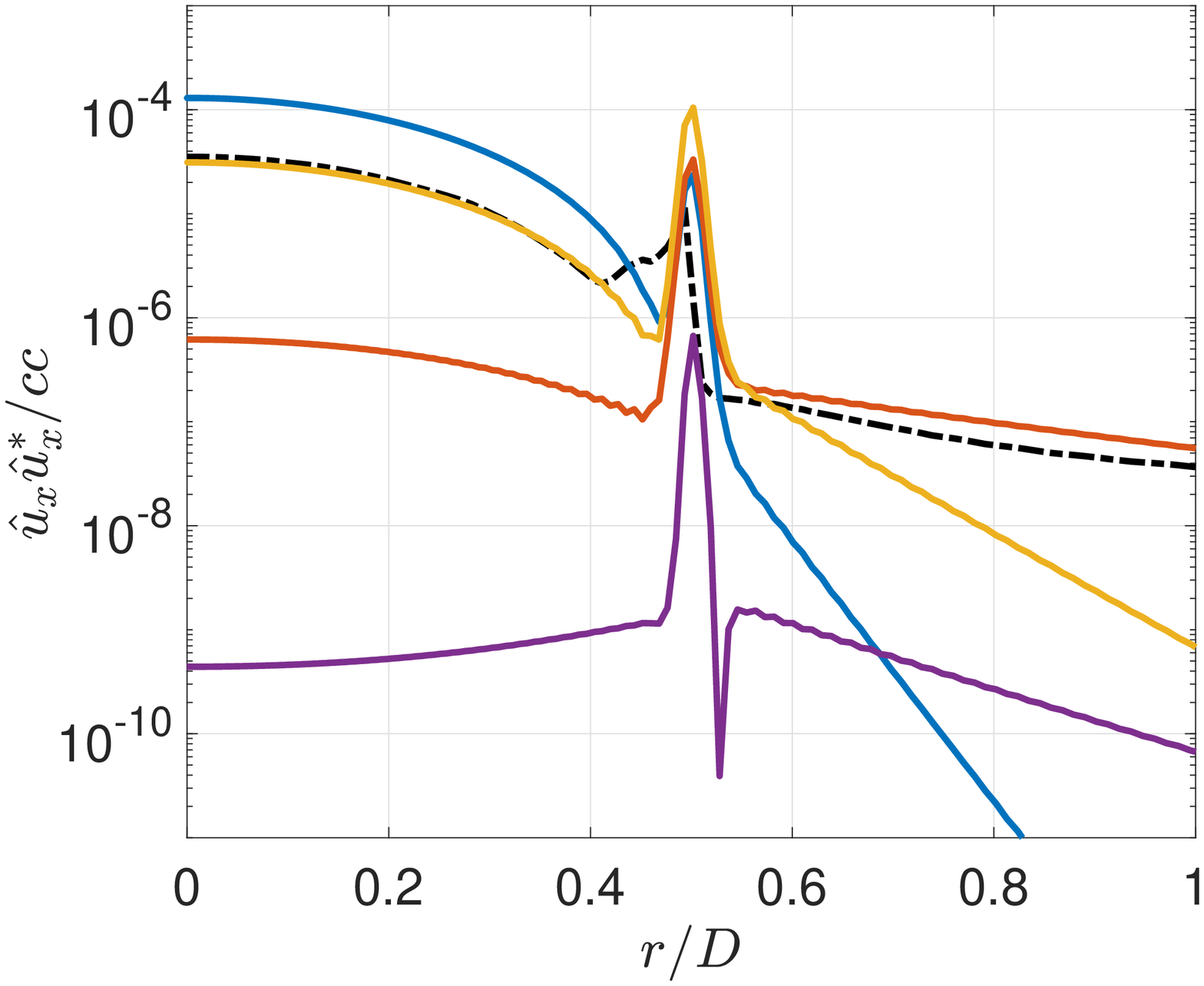}
    \caption{Calibrated $\mathbf{\hat{u}}_x$}
    \label{fig:projections_psd_st03906}
\end{subfigure} 
\begin{subfigure}{0.49\textwidth}
    \includegraphics[width=\textwidth]{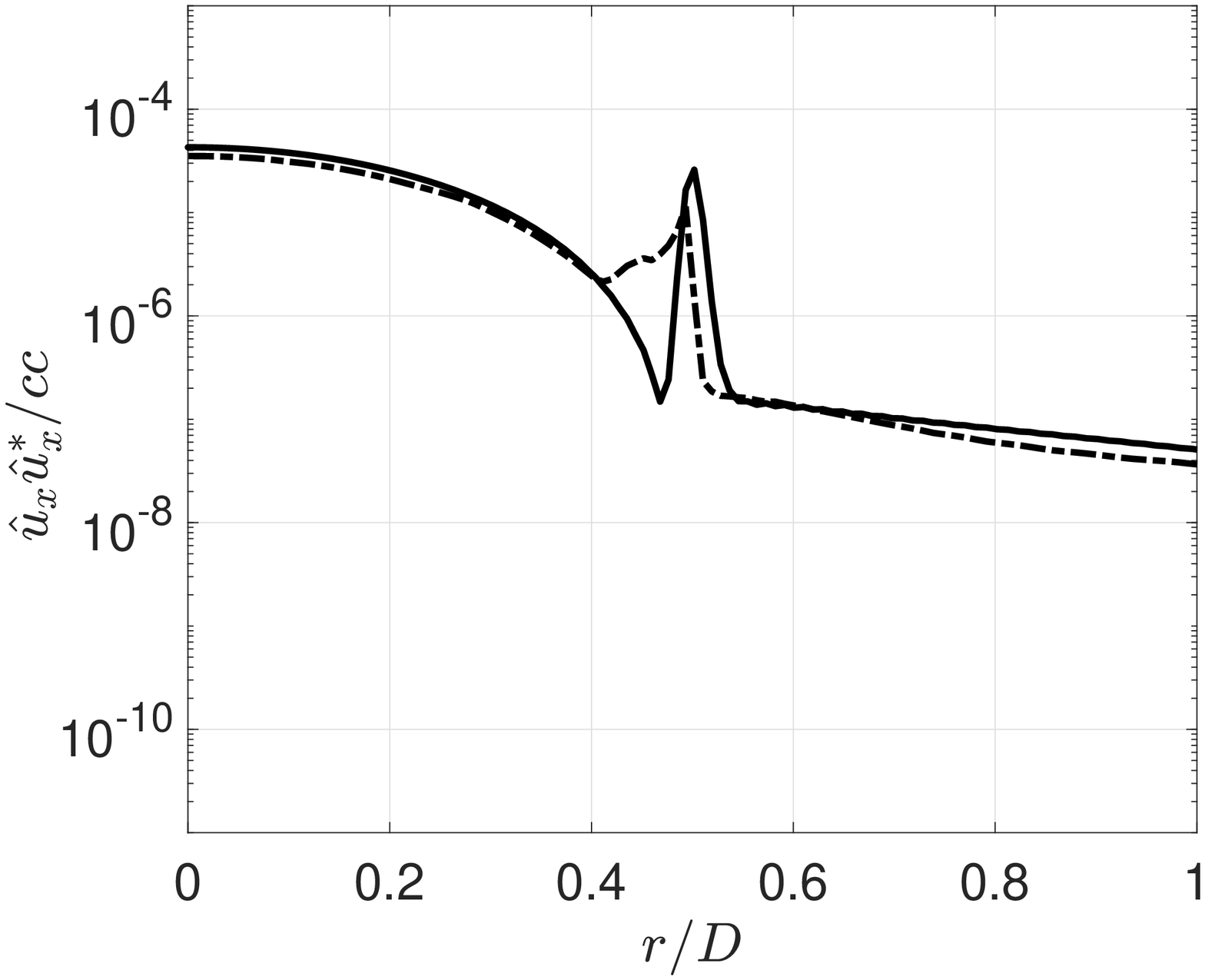}
    \caption{Reconstructed $\mathbf{\hat{u}}_x$}
\label{fig:reconstruction}
\end{subfigure}    
\caption{Projections for $St=0.39$, $m=0$ at $x/D=0$: Streamwise velocity fluctuations PSD (\rule[.45ex]{1.3em}{1pt}, dashed black) LES; (\rule[.45ex]{1.3em}{1pt}, blue) $k^-_{d}$; (\rule[.45ex]{1.3em}{1pt}, red) $k^-_{p}$; (\rule[.45ex]{1.3em}{1pt}, yellow) $k^+_{T}$; (\rule[.45ex]{1.3em}{1pt}, purple) $k^+_{KH}$; (\rule[.45ex]{1.3em}{1pt}, black) reconstruction from 4 modes.}
    \label{fig:projections_reconsturction}
\end{figure}

In figure \ref{fig:projections_reconsturction}(a), radial profiles of the PSDs of the streamwise velocity for a selection of modes are compared with the PSD computed from the LES data for St = 0.39 at the nozzle exit plane.
The guided modes have substantial magnitudes, consistent with the resonance phenomenon observed at this tonal frequency. It is also clear that within the jet core, the $k^-_{d}$ and $k^+_{T}$ dominate the fluctuation field.
In the shear region, $k^-_{d}$, $k^-_{p}$ and $k^+_{T}$ have comparable levels. On the low-speed side of the shear layer, fluctuations are dominated by $k^-_{p}$ and $k^+_{T}$. 
At the nozzle exit plane, negligible KH mode magnitude suggests a rank-3 system locally, but as the KH mode grows exponentially while travelling downstream, the system should be considered rank-4 globally.

A rank-4 reconstruction of the LES data, using these modes, is shown in figure \ref{fig:projections_reconsturction}(b) along with the LES fluctuation profile. At this resonance frequency, the rank-4 model provides a good overall description of the flow dynamics. We see that in the jet core, the reconstruction amplitude is lesser than the amplitude for the most dominant mode (see $k^-_{d}$ in figure \ref{fig:projections_reconsturction}(a)). This is due to the destructive interference between the modes $k^-_{d}$ and $k^+_{T}$ as we found them to be antiphase to each other.
The mismatch for reconstruction in the mixing layer is likely due to disturbances originating in the nozzle boundary layer, leading to energetic but stable mixing layer modes \citep{towne2015one}.
{
These disturbances are insignificant beyond $x/D > 1$ as they decay quickly and do not contribute to the resonance mechanisms, so they need not be taken into account.}

\section{Reflection-coefficient eduction methodology} \label{sec:reflection_method}

We now present a method used to compute reflection-coefficients between pairs of $k^-$ and $k^+$ waves at the resonance end locations.
Referring back to figure \ref{fig:nozzle}, at the turning point, an incident $k^+_T$ wave reflects as a $k^-_d$ or $k^-_p$ wave and transmits as an evanescent wave. 
The reflected wave propagates upstream to the nozzle exit, where it reflects as a $k^+_T$ wave, which then propagates downstream to the turning point, completing the resonance loop.
The relation of magnitude and phases of these waves among each other are described by reflection and transmission coefficients at the corresponding end locations. Note that at the nozzle plane, apart from the contribution from $k^-_d$ or $k^-_p$ waves, $k^+_T$ wave may also be driven by nozzle fluctuations, or by the reflection of other upstream-travelling $k^-$ waves.

\subsection{Reflection equations for nozzle exit plane} \label{subsec:reflection_eq_x0}

At the nozzle exit plane, the expansion coefficients of the $k^+_{T}$ wave are related to expansion coefficients of $k^-$ waves through complex reflection-coefficients as
    \begin{equation} \label{eq:reflection_eq1}
   {a}_{T}^+ = {R_{n,d-}} \hspace{1mm} {a}_{d}^- +  {R_{n,p-}} \hspace{1mm} {a}_{p}^- +  {a}_{o}
 \end{equation}
where, {${R_{n,d-}}$ is the reflection-coefficient for $k^{-}_d$  reflecting into $k^{+}_T$ and 
${R_{n,p-}}$ is the reflection-coefficient for  $k^{-}_p$ reflecting into $k^{+}_T$.
 }
Here, ${R_{n,d-}} \hspace{1mm} {a}_{d}^-$ is the 
contribution to ${a}_{T}^+$ that arises from reflection of $   {a}_{d}^-$;
${R_{n,p-}} \hspace{1mm} {a}_{p}^-$ is the contribution from reflection of ${a}_{p}^-$;
$ {a}_{o}$ groups all other contributions, e.g., reflections of other waves or disturbances coming from within the nozzle.

To evaluate the reflection-coefficients ${R_{n,d-}} $ and ${R_{n,p-}}$, following the procedure of \citet{bendat2011random}, we multiply \eqref{eq:reflection_eq1} with both ${{a}_{d}^-}^H$ and ${{a}_{p}^-}^H$ and take the expected value, giving 
    \begin{align} \label{eq:reflection_eq3}
  \langle {a}_{T}^+  {{a}_{d}^-}^H \rangle =  {R_{n,d-}} \langle {a}_{d}^-  {{a}_{d}^-}^H \rangle  +   {R_{n,p-}} \langle {a}_{p}^-  {{a}_{d}^-}^H \rangle  +   \langle {a}_{o} {{a}_{d}^-}^H \rangle , \\
 \label{eq:reflection_eq5}
 \langle {a}_{T}^+  {{a}_{p}^-}^H \rangle =  {R_{n,d-}} \langle {a}_{d}^-  {{a}_{p}^-}^H \rangle  +   {R_{n,p-}} \langle {a}_{p}^-  {{a}_{p}^-}^H \rangle  +   \langle {a}_{o} {{a}_{p}^-}^H \rangle . 
 \end{align}

We assume that the contributions from the nozzle boundary layer disturbances and other wave reflections are uncorrelated with the resonance dynamics, thus $\langle {a}_{o} {{a}_{d}^-}^H \rangle = \langle {a}_{o} {{a}_{p}^-}^H \rangle = 0 $. With this assumption, \eqref{eq:reflection_eq3} and \eqref{eq:reflection_eq5} can be solved, expressing the reflection-coefficients, ${R_{n,d-}}$ and ${R_{n,p-}}$, in terms of expansion coefficient correlations.

%
  
  \subsection{Reflection equations for turning point plane} \label{subsec:reflection_eq_xTP}
  
  We now present the system of equations used to calculate the reflection-coefficients at the turning point location where the $k^+_{T}$ reflects as $k^-_{d}$ and $k^-_{p}$ (figure \ref{fig:nozzle}). 
   Turning point locations are educed from the saddle point curves presented in figure \ref{fig:resonance_mechanism1}.
   
   Following a similar procedure to that of section \ref{subsec:reflection_eq_x0}, we can say that, at the turning point,    
    \begin{equation} \label{eq:turning_reflection_eq1}
    \internallinenumbers
   {a}_{d}^- = {R_{tp,d-}} \hspace{1mm} {a}_{T}^+ + {a}_{o}  \hspace{1cm} \&  \hspace{1cm}   {a}_{p}^- = {R_{tp,p-}} \hspace{1mm} {a}_{T}^+ + {a}_{o},  \\
 \end{equation}
 where ${R_{tp,d-}}$ and ${R_{tp,p-}}$ are the turning point reflection-coefficients  {for $k^{+}_T$ reflecting into $k^{-}_d$ and $k^{-}_p$ respectively.}

Multiplying  \eqref{eq:turning_reflection_eq1} with 
${{a}_{T}^+}^H$ and taking the expected value gives, 
     \begin{equation} \label{eq:turning_reflection_eq2}
 {\langle {a}_{d}^-  {{a}_{T}^+}^H  \rangle} ={R_{tp,d-}} {\langle {a}_{T}^+  {{a}_{T}^+}^H \rangle}+\langle {a}_{o} {{a}_{T}^+}^H \rangle \hspace{0.3cm} \&  \hspace{0.3cm}   {\langle {a}_{p}^-  {{a}_{T}^+}^H \rangle}={R_{tp,p-}}{\langle {a}_{T}^+  {{a}_{T}^+}^H \rangle} +\langle {a}_{o} {{a}_{T}^+}^H \rangle.
 \end{equation}
With the assumption of $\langle {a}_{o} {{a}_{T}^+}^H \rangle = 0$, \eqref{eq:turning_reflection_eq2} can be solved for ${R_{tp,d-}}$ and ${R_{tp,p-}}$.

  \section{Results and discussions} \label{sec:results}
  \subsection{Coherence analysis} \label{subsec:coh_analysis}

Before evaluating the reflection-coefficients, we examine the relation between expansion coefficient signals from the modes through the coherence function, 
\begin{equation} \label{eq:coherence_function}
\gamma^2_{{12}} = \frac{ \langle {a}_{{{1}}} {a}_{{2}}^H \rangle^2} {\langle {a}_{{1}} {a}_{{1}}^H \rangle \hspace{1mm} \langle {a}_{{2}} {a}_{{2}}^H \rangle} ,
\end{equation}

\begin{figure} 
    \centering
\begin{subfigure}{0.49\textwidth}
    \includegraphics[width=1\textwidth]{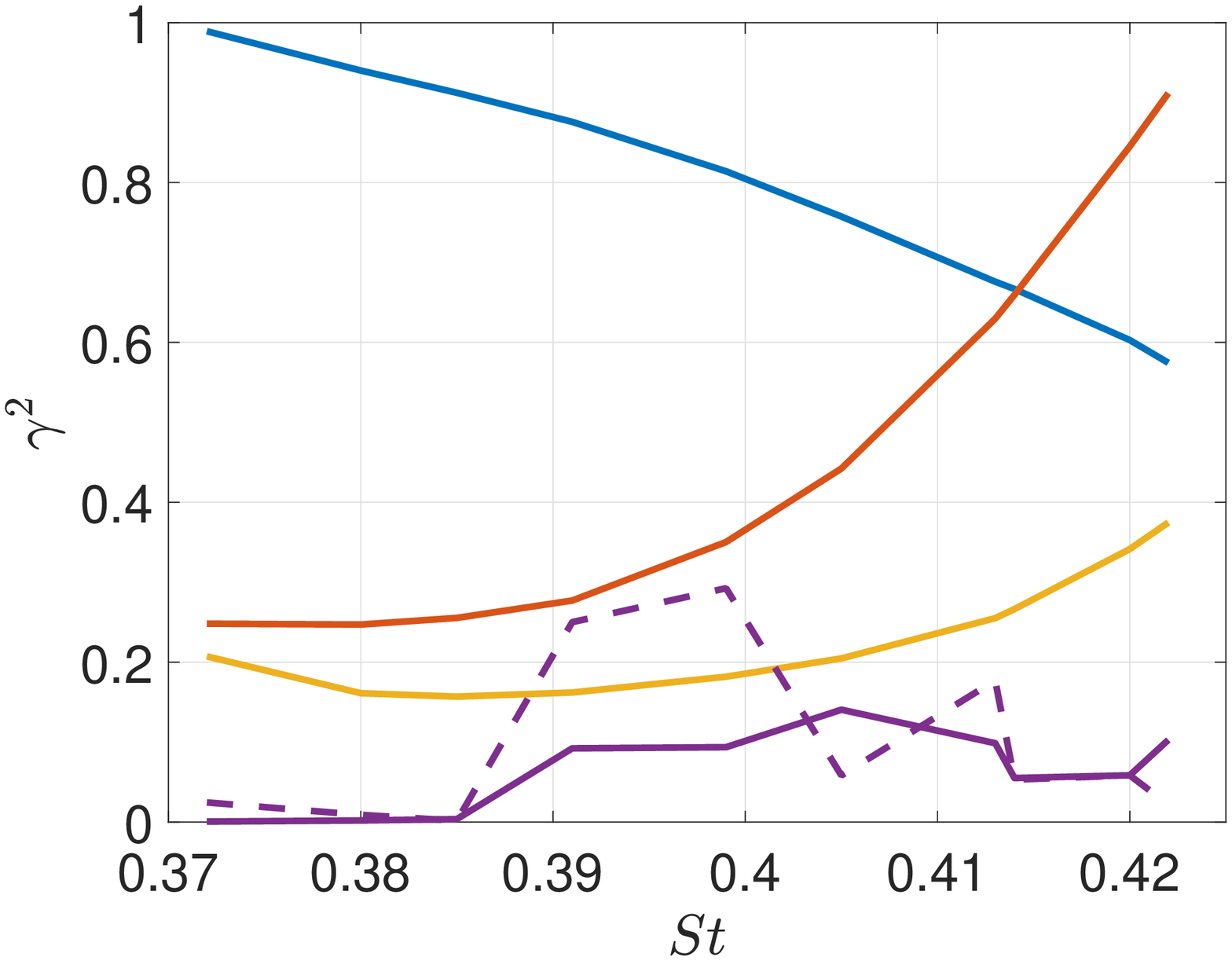}
    \caption{Coherence function}
    \label{fig:coherence_vs_St}
\end{subfigure} 
\begin{subfigure}{0.49\textwidth}
    \includegraphics[width=\textwidth]{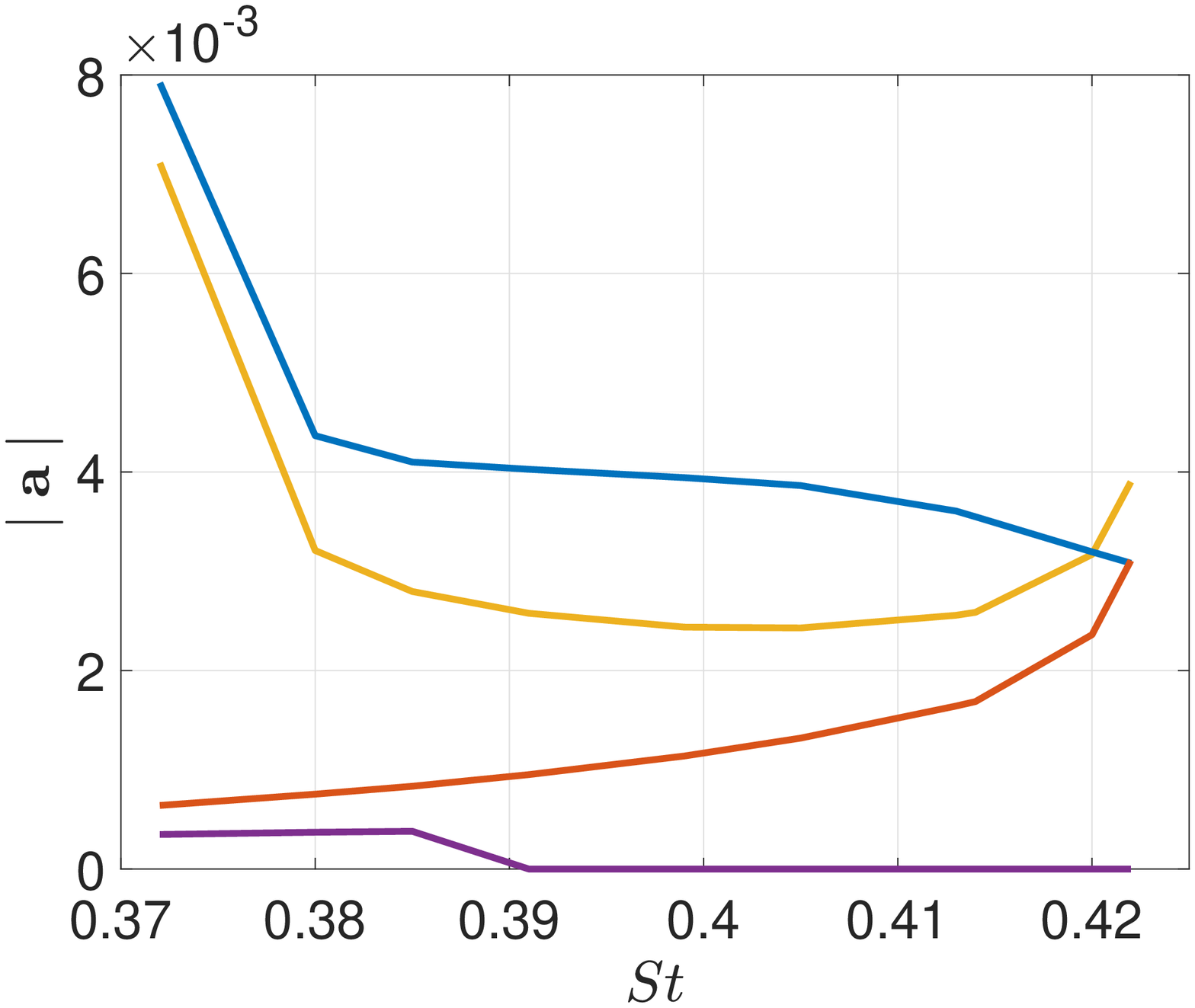}
    \caption{Mode amplitudes}
\label{fig:BOPAmplitudes_vs_St}
\end{subfigure}    
\caption{At the nozzle exit, $x/D=0$. (a) Coherence function: (\rule[.45ex]{1.3em}{1pt}, blue) $k_d^-/k_T^+$; (\rule[.45ex]{1.3em}{1pt}, red) $k_p^-/k_T^+$; (\rule[.45ex]{1.3em}{1pt}, yellow) $k_d^-/k_p^-$; (\rule[.45ex]{1.3em}{1pt}, purple) $k_d^-/k_{KH}^+$; (\rule[.45ex]{1.3em}{1pt}, dashed purple) $k_p^-/k_{KH}^+$.
(b) Mode amplitudes: (\rule[.45ex]{1.3em}{1pt}, blue) $k^-_{d}$; (\rule[.45ex]{1.3em}{1pt}, red) $k^-_{p}$; (\rule[.45ex]{1.3em}{1pt}, yellow) $k^+_{T}$; (\rule[.45ex]{1.3em}{1pt}, purple) $k^+_{KH}$.}
\label{fig:coherence_BopAmplitudes}
\end{figure}

\noindent where ${a}_{{1}}$ and ${a}_{{2}}$ are the expansion coefficients (see \eqref{eq:biorthogonal_projection}) for the two modes, and $\langle\cdot\rangle$ represents the expected value, derived from the available realisations, obtained by averaging across these realisations.
{This relation provides an initial insight into the resonance mechanisms at play. A pair of \( k^+ \) and \( k^- \) modes exhibiting strong coherence magnitudes suggests that they are the resonating pair of modes, reflecting into each other at the resonance end locations.}

For the present system of modes, coherence-function dependence on $St$ at $x/D=0$ can be seen in figure \ref{fig:coherence_BopAmplitudes}(a). 
For low $St$,  a strong coherence is observed between $k^-_{d}$ and $k_{T}^+$ which suggests that the $k^-_{d}/k_{T}^+$ resonance pair is active at these frequencies.

As $St$ increases, the coherence function decays for $k^-_{d}/k^+_{T}$ while rises sharply for $k^-_{p}/k^+_{T}$. This suggests a change in the resonance mechanism, as frequency increases, towards a scenario where the $k^-_{p}/k^+_{T}$ pair is dominant. The small coherence of the $k^+_{KH}$ mode ($\gamma^2<0.3$) with the $k^-$ waves signifies its absence in the resonance mechanisms, and for this reason, it is excluded from the forthcoming discussion.

The variation of mode amplitudes with $St$ at $x/D=0$, as shown in figure 
\ref{fig:coherence_BopAmplitudes}(b), tells a similar story. At low $St$, the $k_T^+$ amplitude decays with increasing $St$ following the trend of $k_d^-$; but at high $St$, it grows with increasing $St$, following the trend of $k_p^-$, again reflecting a change of the dominant resonant mechanisms with increasing frequency.

  \subsection{Reflection-coefficients at the resonance end locations} \label{sec:results_RC}

\begin{figure}
    \centering
\begin{subfigure}{0.49\textwidth}
    \includegraphics[width=1\textwidth]{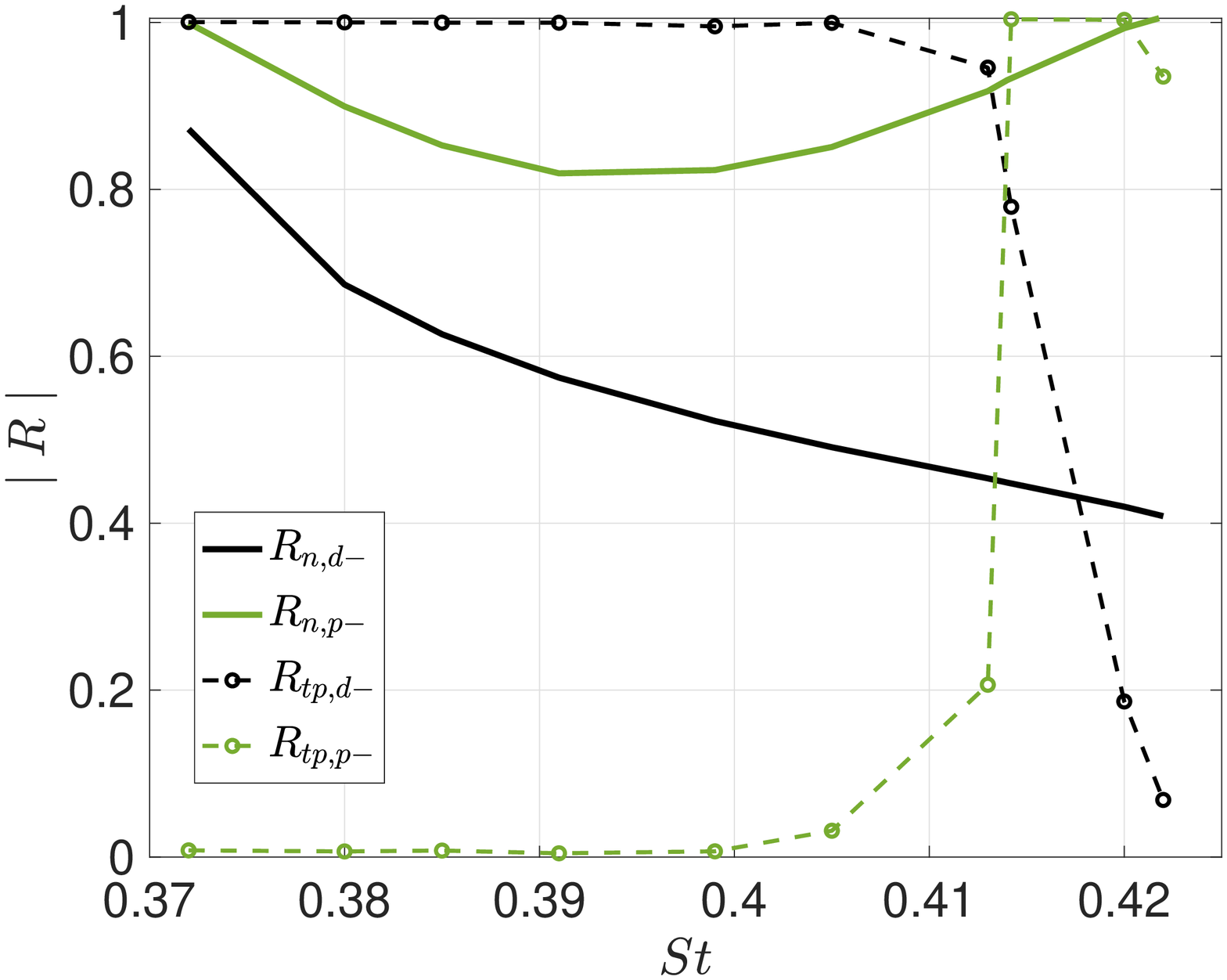}
    \caption{Reflection-coefficient magnitude}
    \label{fig:ref_mag}
\end{subfigure} 
\begin{subfigure}{0.49\textwidth}
    \includegraphics[width=\textwidth]{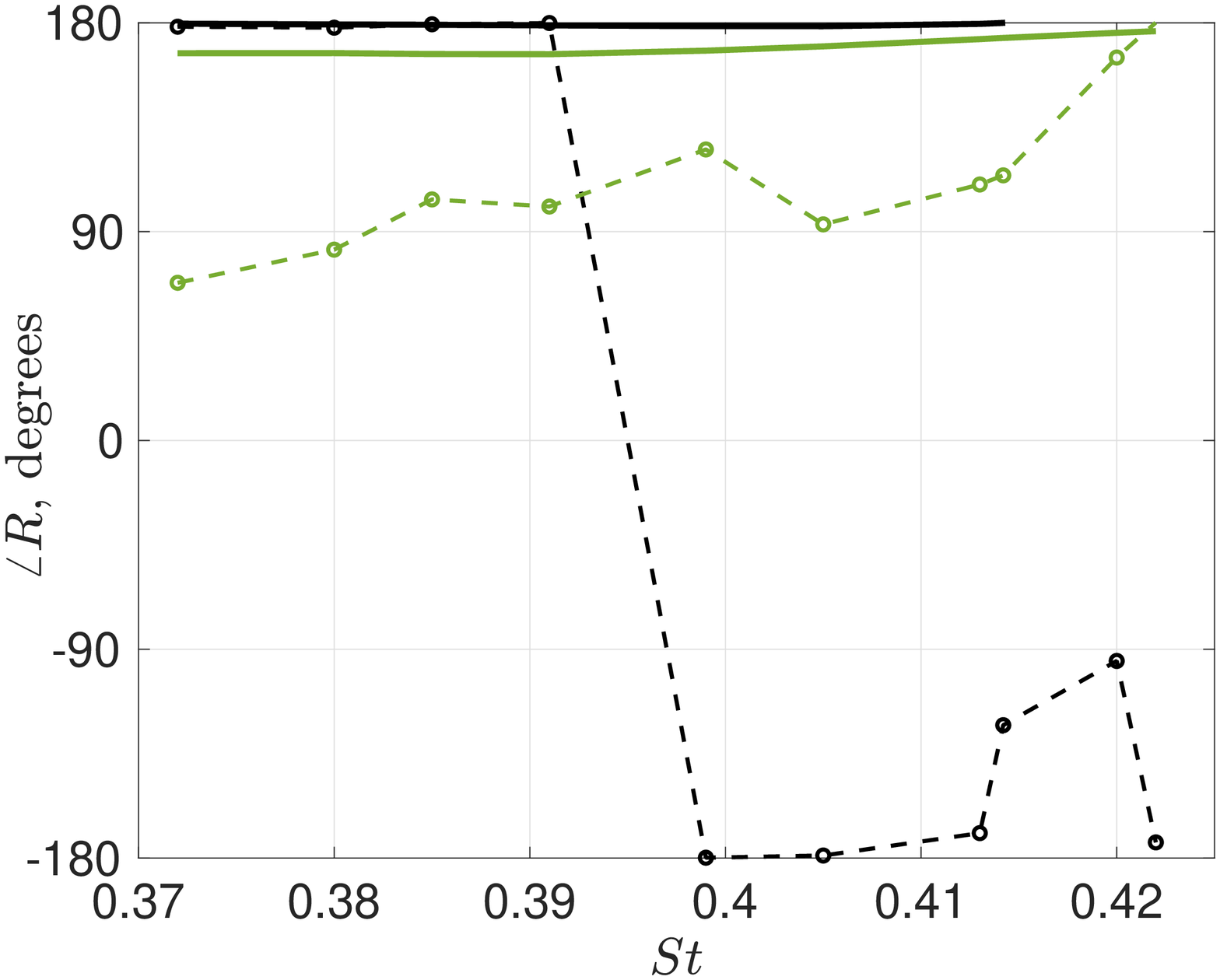}
    \caption{Reflection-coefficient phase}
\label{fig:ref_phase}
\end{subfigure}    
\caption{Reflection-coefficients. At the nozzle exit: ${R_{n,d-}}$ and ${R_{n,p-}}$ correspond to the reflection of $k_d^-$ wave and $k_p^-$ wave respectively into the $k_T^+$ wave. At the turning point: ${R_{tp,d-}}$ and ${R_{tp,p-}}$ correspond to the reflection of $k_T^+$  wave into the $k_d^-$  wave and $k_p^-$ wave respectively.}
\label{fig:refcoeff}
\end{figure}

In the nozzle exit plane, the magnitudes and phases of the reflection-coefficients,  ${R_{n,d-}}$ and ${R_{n,p-}}$,  are presented as a function of $St$ in figures \ref{fig:refcoeff}(a) and \ref{fig:refcoeff}(b). High magnitudes for both the reflection-coefficients indicate strong reflections. We also observe that as $St$ increases, $\mid {R_{n,d-}}\mid $ decreases while $\mid  {R_{n,p-}}\mid $ decreases and then increases. 
The phase angles for both reflection-coefficients are close to $180^o$ indicating out-of-phase reflection.

At the turning point, the magnitudes and phases of the reflection-coefficients are shown in figures \ref{fig:refcoeff}(a) and \ref{fig:refcoeff}(b) as well. 
From the local stability analysis, the $k_T^+$ mode is evanescent downstream of the turning point (see figure \ref{fig:eigenspectrum_trajectories}(b)). This implies perfect reflection in the turning-point plane, as beyond here, the $k_T^+$ mode cannot propagate energy downstream. This is exactly what is found for the reflection-coefficients in the turning-point plane i.e. $\mid {R_{tp,d-}}\mid \sim 1$ \& $\angle {R_{tp,d-}} \sim 180^o$ for the lower $St$ ($0.37<St<0.41$); and $\mid {R_{tp,p-}}\mid \sim 1$ \& $90<\angle {R_{tp,p-}} < 180^o$ for the higher $St$ ($0.41<St<0.43$).

\subsection{Resonance-mechanism dependence on $St$} \label{sec:results_resonanceMechanisms}
  The results from sections \ref{subsec:coh_analysis} and \ref{sec:results_RC} conclude that 
for the frequency range $0.37<St<0.41$, it is the $k_d^-/k_T^+$ pair of modes that resonates while for
$0.41<St<0.43$, it is the $k_p^-/k_T^+$ pair that resonates. The resonance mechanism switches near $St=0.415$. 

These two resonance mechanisms were proposed by \citet{towne2017acoustic} and can be seen in figure  \ref{fig:resonance_mechanism1}.
For the lower frequencies, $St=0.39$ for instance, the saddle point at the turning point exists between $k^-_d $ and $ k^+_T$ modes, which means that the acoustic resonance at $St=0.39$ is being led by the $k^+_{T}/k^-_{d}$ pair. This is exactly what we see in figures \ref{fig:coherence_BopAmplitudes} and \ref{fig:refcoeff}(a), where the $k^-_d / k^+_T$ pair of modes has strong coherence and large reflection-coefficient magnitudes.
  
  \begin{figure}
    \centering
\begin{subfigure}{0.49\textwidth}
    \includegraphics[width=1\textwidth]{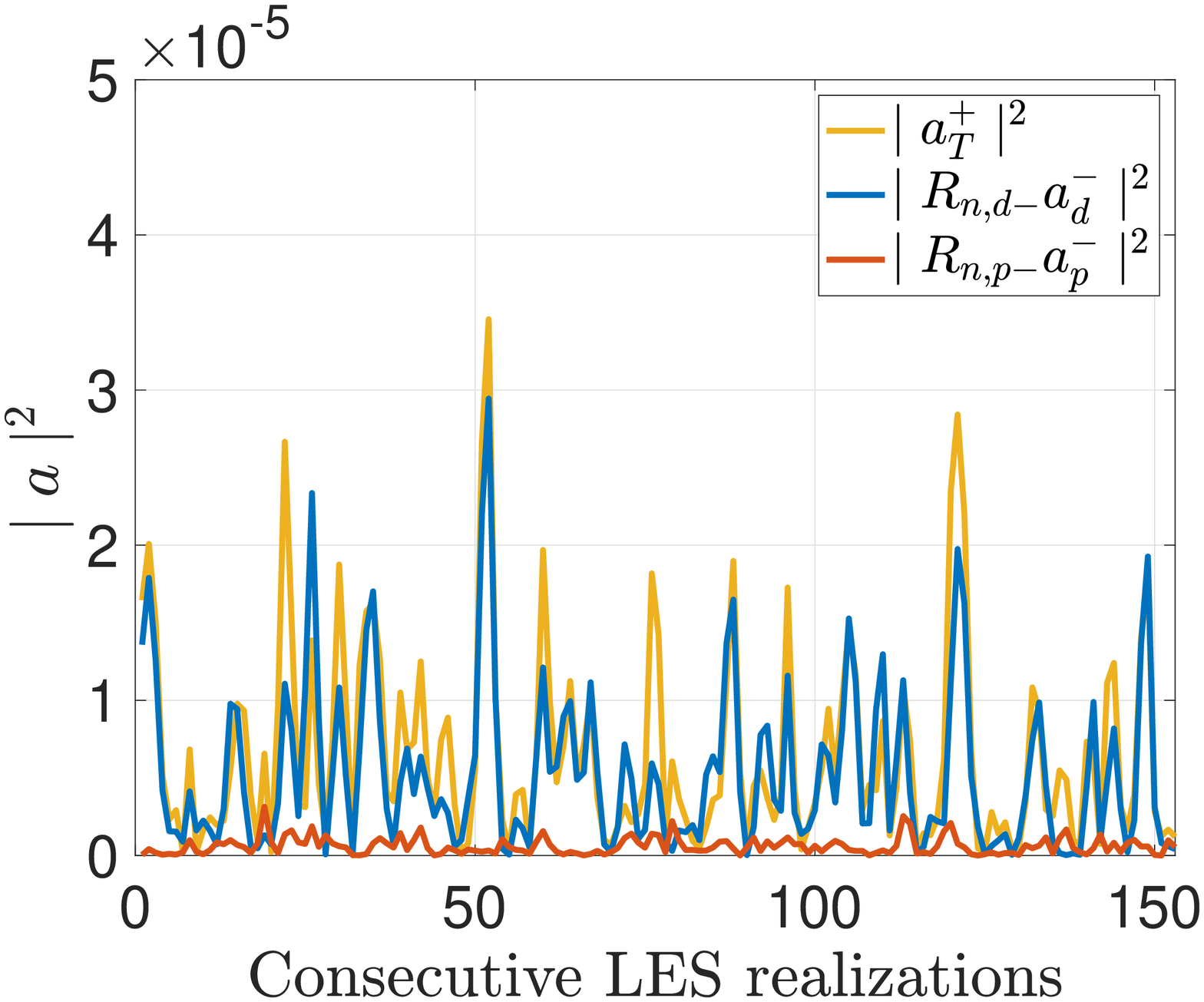}
    \caption{For $St=0.39$}
    \label{fig:contribution_st039_x00}
\end{subfigure}  
\begin{subfigure}{0.49\textwidth}
    \includegraphics[width=1\textwidth]{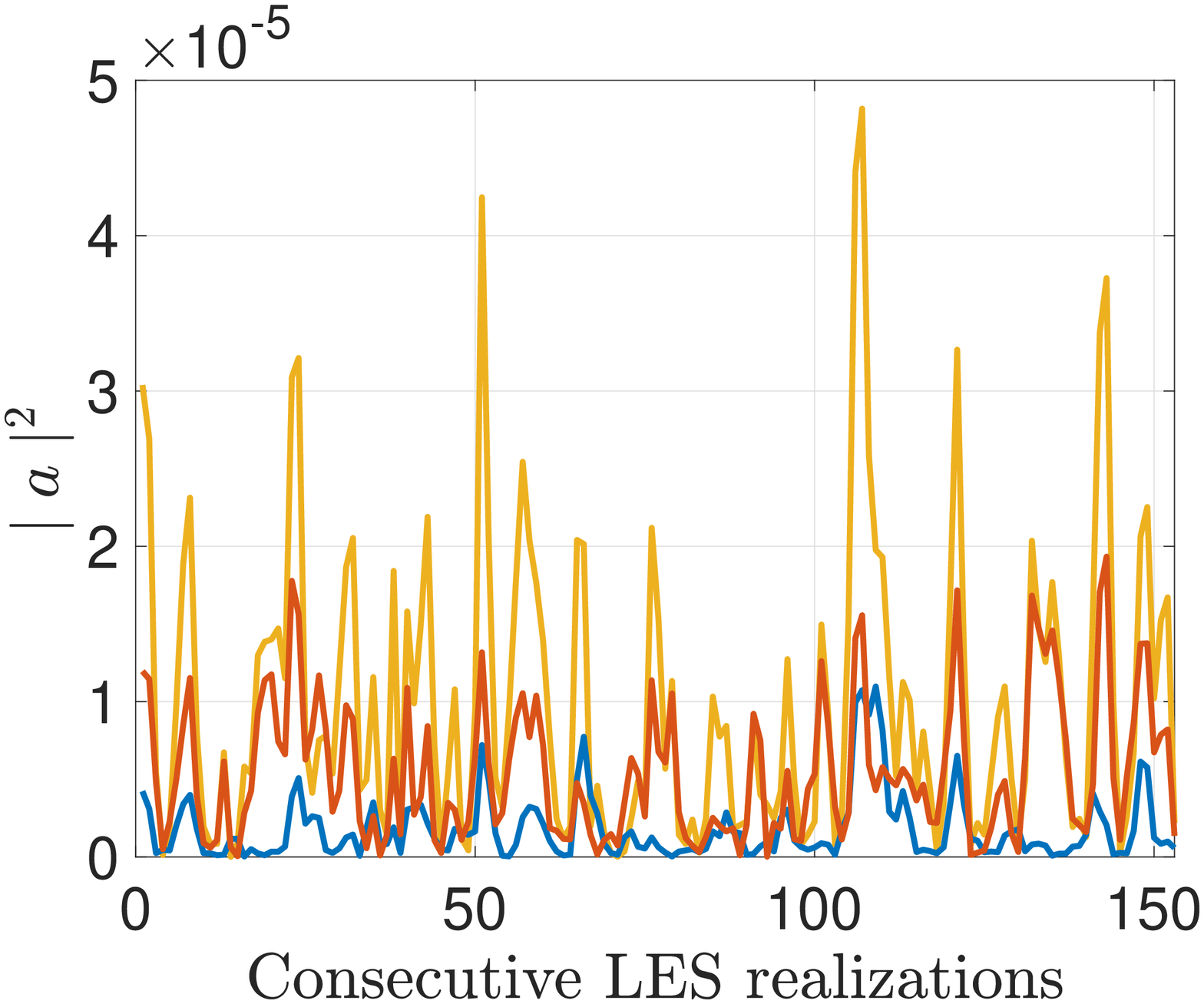}
    \caption{For $St=0.42$ }
    \label{fig:contribution_st042_x00}
\end{subfigure}  
\caption{Contribution of $k^-_{d}$ and $k^-_{p}$ to $k^+_{T}$ at the nozzle exit plane.}
\label{fig:new}
\end{figure}

    For the reflections at the nozzle-exit plane at $St=0.39$, the individual contributions of $k^-_{d}$ and $k^-_{p}$ to $k^+_{T}$ are better seen in figure \ref{fig:new}(a) (see  \eqref{eq:reflection_eq1} for reference).
 The plot displays the square magnitude of mode amplitudes for $k^+_{T}$ (in yellow), as well as the contributions of $k^-_{d}$ and $k^-_{p}$ (in blue and red, respectively) across consecutive LES realizations. 
Figure \ref{fig:new}(a) demonstrates that $k^+_{T}$ is primarily underpinned by reflection of $k^-_{d}$. Despite the high magnitude of ${R_{n,p-}}$ at the nozzle-exit plane (figure \ref{fig:refcoeff}(a)), the  contribution of the  $k^-_{p}$ is  much smaller than that of  $k^-_{d}$, due its smaller amplitude (figure \ref{fig:coherence_BopAmplitudes}(b)).
  
    For the higher frequencies, $St=0.42$ for instance, the saddle point at the turning point exists between $k^-_p $ and $ k^+_T$ modes, hence the acoustic resonance is governed by the $k^+_{T}/k^-_{p}$ pair. This is also what we observe in figures \ref{fig:coherence_BopAmplitudes} and \ref{fig:refcoeff}(a). 
    The individual contributions of $k^-_{d}$ and $k^-_{p}$ to $k^+_{T}$ in the figure \ref{fig:new}(b) shows that $k^+_{T}$ follows $k^-_{p}$ much more closely than $k^-_{d}$ at this $St$. Hence, $k^+_{T}$ is the direct reflection result of $k^-_{p}$ at $St=0.42$.

\section{Conclusion} \label{sec:conclusions}
Resonating guided waves in the potential core of a Mach 0.9 turbulent jet which lead to tones previously observed in experiments and numerical simulations \citep{towne2017acoustic,bres2018importance} have been studied. 
The resonating guided waves consisted of a downstream-travelling duct-like wave ($k_T^+$), an upstream-travelling duct-like wave  ($k_d^-$), and an upstream-travelling {discrete shear-layer}  wave  ($k_p^-$). 
 Through a comprehensive analysis of time-resolved turbulent jet data,   the presence of these frequency-dependent resonance phenomena, within the specified tonal frequency range, has been demonstrated.

The reflection-coefficients associated with energy exchange at the resonance end locations have also been evaluated. Such evaluations are crucial for refining simplified resonance models, such as by \citet{jordan2018jet, mancinelli2019screech, mancinelli2021complex}, {in which the reflection-coefficients, being unknown, amount to parameters by which a model can be tuned to match data, rather than be informed a priori based on flow physics. The present method provides a physics-based estimate of these reflection coefficients, which may be  used both as input for such resonance models, and, later, to support the elaboration of models of the reflection  process in itself.  }

Bi-orthogonal projection of LES data onto eigenmodes obtained from a linear stability analysis based on the turbulent mean flow was used to provide amplitudes of the resonating waves at the resonance end locations: the nozzle exit plane and downstream turning points. 
The dynamics of the flow at resonance frequencies are well described by a rank-4 model, comprising these neutrally stable guided waves ($k_T^+$, $k_p^-$ and $k_p^-$) and K-H instability wave.
The reflection-coefficients at the resonance end locations were computed under the assumption that contributions from non-resonant modes are uncorrelated with the resonant modes. For the range of tonal frequencies, $0.37<St<0.43$, the mode amplitudes, coherence among them, and reflection-coefficients were presented. 

Depending on the frequency, either of the $k^-$ waves was found to be taking part in the resonance loop i.e. for $0.37<St<0.41$ (F1 frequency range), the pair $k_d^-/k_T^+$ was active while for $0.41<St<0.43$ (F2 frequency range), the pair $k_p^-/k_T^+$ was active. 
This frequency-dependence of resonance mechanism had been suggested by \citet{towne2017acoustic} 
where it was shown that the $k_T^+$ mode forms a turning point at the saddle point where upstream- and downstream-travelling waves exchange energy, with $k_d^-$ mode in the F1 frequency range and with $k_p^-$ mode in the F2 frequency range.
{Among the five possible resonance mechanisms postulated by \citet{towne2017acoustic}, these two have now been identified as those that underpin the peaks observed in the experimental and LES data.}

Although this study focuses on a specific case of an isothermal Mach 0.9 jet, the methodologies and analyses presented can be applied to other scenarios such as: 
(i) Jet-flap interaction noise, where resonance end location at the downstream is the flap, a physical boundary and the downstream-traveling $k^+$ mode involved in resonance is the K-H mode \citep{jordan2018jet}; (ii) Screech tones in supersonic jets, where shock cells act as the resonance end locations \citep{edgington2019aeroacoustic, mancinelli2019screech}; among other potential applications.

\backmatter
\bibliography{references} 

\end{document}